\documentclass[natbib,twocolumn,twoside]{svmultiag}
\usepackage[dvips]{graphicx}
\graphicspath{{./figures/}}
\newcommand{\psrpi}{{PSR\large$\pi$}}
\newcommand{\uas}{\ensuremath{\mu\mathrm{as}}}

\newcommand{\degrees}{\ensuremath{^\circ}}

\title*{PSR{\boldmath\LARGE$\pi$\unboldmath}: A large VLBA pulsar astrometry program}
\titlerunning{PSR{\normalsize$\pi$}: A large VLBA pulsar astrometry program}

\author{A.~T.~Deller \and W.~F.~Brisken \and S.~Chatterjee \and J.~M.~Cordes \and W.~M.~Goss \and G.~H.~Janssen \and Y.~Y.~Kovalev \and T.~J.~W.~Lazio \and L.~Petrov \and B.~W.~Stappers}
\authorrunning{Deller et al.}

\institute{A.~T.~Deller \at ASTRON, Oude Hoogeveensedijk 4,
7991 PD Dwingeloo,
The Netherlands
\and 
W.~F.~Brisken
\and
W.~M.~Goss \at National Radio Astronomy Observatory, Socorro, NM 87801, USA
\and
S.~Chatterjee
\and
J.~M.~Cordes \at Astronomy Dpt., Cornell University, Ithaca, NY 14853, USA
\and
G.~H.~Janssen
\and
B.~W.~Stappers \at Jodrell Bank Centre for Astrophysics, School of Physics and Astronomy, University of Manchester, Manchester M13 9PL, UK
\and
Y.~Y.~Kovalev \at Astro Space Center of Lebedev Physical Institute, Profsoyuznaya 84/32, Moscow 117997
\and
T.~J.~W.~Lazio \at Jet Propulsion Laboratory, California Institute of Technology, 4800 Oak Grove Dr., Pasadena, CA 91109, USA
\and
L.~Petrov \at ADNET Systems Inc./NASA GSFC, Greenbelt, MD 20771, USA
          }
\begin{document}
\maketitle
\abstract{
Obtaining pulsar parallaxes via relative astrometry (also known as differential astrometry) yields distances and transverse velocities that can be used to probe properties of the pulsar population and the interstellar medium. Large programs are essential to obtain the sample sizes necessary for these population studies, but they must be efficiently conducted to avoid requiring an infeasible amount of observing time. This paper describes the \psrpi\ astrometric program, including the use of new features in the DiFX software correlator to efficiently locate calibrator sources, selection and observing strategies for a sample of 60 pulsars, initial results, and likely science outcomes.  Potential applications of high-precision relative astrometry to measure source structure evolution in defining sources of the International Celestial Referent Frame are also discussed.
}


\section{Introduction}                                \label{sec:introduction}
Due to their unique combination of high density, high magnetic field and high angular momentum,
pulsars provide a rich laboratory for investigating phenomena in the fields of nuclear physics, 
particle physics, gravitational physics and many others \citep[see e.g.,][and references therein]{lorimer08a}.
However, studies of pulsars are hampered by the uncertainty in distance--dependent quantities 
introduced by the reliance on dispersion measure (DM) distance estimates.  Due to the 
highly non--uniform distribution of ionized material in the ISM on sub-kpc scales, 
the correspondence between a pulsar's DM and its distance often remains uncertain, despite the 
development of detailed models of the ionized interstellar medium (e.g., the TC93 model;
\citealt{taylor93a}, and the NE2001 model; \citealt{cordes02a}).  Although distances estimated 
from the TC93 and NE2001 models are generally assumed to be accurate to within
20\%, previous astrometric pulsar observations have shown that much greater errors are
sometimes possible for individual objects 
\citep[e.g., a factor of 4.5 error in the distance for PSR~J0630--2834;][]{deller09b}, and systematic 
biases are likely \citep{lorimer06a}.

Thus, independent distance measures to pulsars are vital, both to enable the confident 
estimation of distance--dependent parameters for individual pulsars and to refine
DM--based distance models for the remainder of the pulsar population. 
Whilst distance estimates can also be provided from associations with other astrophysical objects \citep[e.g.][]{camilo06a} 
or annual geometric parallax measurements made via timing \citep[e.g.,][]{hotan06a}, parallax
measurement using Very Long Baseline Interferometry \citep[VLBI; e.g.,][]{chatterjee09a} 
is the most widely applicable technique and has provided the most accurate 
distance measurements \citep{deller08b}.

The \psrpi\ program is using the
Very Long Baseline Array (VLBA\footnote{The VLBA is operated by the National Radio Astronomy Observatory as a facility of 
the National Science Foundation, operated under cooperative agreement by Associated Universities, Inc.})
to undertake a large pulsar astrometry program, taking advantage of the VLBA's 
relatively large field of view to observe ``in--beam" calibrators for the best astrometric accuracy.
Taking advantage of this capability requires the identification of suitable calibrator sources,
which was the first (now largely complete) phase of the \psrpi\ project.  
The second phase, which is underway now,
involves astrometric observations for 60 pulsars spread over a 1.5 year period.  A third
phase, not yet approved, would expand the program to 200 pulsars once the 
ongoing VLBA sensitivity upgrade is complete\footnote{The VLBA 
sensitivity upgrade program is described at http://www.vlba.nrao.edu/memos/sensi/}.

In this paper, preliminary results of the \psrpi\ program are presented, including
the success of the in--beam calibrator search program and the verification of
identified calibrators in initial astrometric epochs.  The current status of the program
and plans for the next 18 months are presented, along with the expected program
outcomes.  Finally, new techniques for and applications of high precision astrometric datasets
are discussed, including the potential for direct
and accurate measurement of $\mu$as--level variation in the structure of
sources which define the International Celestial
Reference Frame (ICRF).

\section{In--beam calibrator search observations}   \label{sec:inbeam}
The steep spectrum exhibited by most pulsars dictates that a large astrometric survey 
using current VLBI facilities must
observe at relatively low frequency ($\sim$1.6 GHz) where enough pulsars are
sufficiently bright.  At these frequencies, the predominant source of systematic
contribution to astrometric error is the differential ionosphere between the calibrator
and target sources, and so minimising the calibrator--target angular separation 
is the most important consideration for obtaining accurate astrometry \citep[e.g.,][]{chatterjee04a}.  An ``in--beam"
calibrator, which can be observed contemporaneously with the pulsar target, is
always preferred.  However, on average the distance to a known VLBI calibrator source
is $\sim$2\degrees, much greater than the primary beam width of a 25m antenna
(such as is used in the VLBA).  Thus, suitably bright and nearby compact calibrators must be 
identified prior to the astrometric program.  The 1$\sigma$\ sensitivity
of a single VLBA baseline at 1.6 GHz is 1.7 mJy in 5 minutes at the current maximum bandwidth of
128 MHz; this will improve to 0.8 mJy at the 512 MHz bandwidth to become 
available in 2011.  Thus, if a S/N ratio of 10 is required at each of the VLBA's 10 stations, 
calibrators of peak flux density $\ge$6 mJy can be used currently, while those $\ge$3 mJy will 
become available with future higher bandwidths.

In the past, this ``in--beam search" operation was time--consuming, due to the
need to identify sources likely to be compact before snapshot VLBI observations were used to 
determine their true flux one at a time.  However, a new ``multifield" mode of the DiFX software
correlator \citep{deller11a} enables all known sources within the telescope primary beam
to be inspected at VLBI resolution simultaneously, meaning only a single snapshot 
observation is required, regardless of the number of candidates.  The VLBA can 
reach a 1$\sigma$ sensitivity of 300 $\mu$Jy in less than 4 minutes on--source,
and so phase--referenced VLBI observations can quickly and reliably detect all 
VLBA in--beam calibrators that will be useful at current or future data rates.

\begin{figure}[tb]
         \centering
         \includegraphics[width=.5\textwidth]{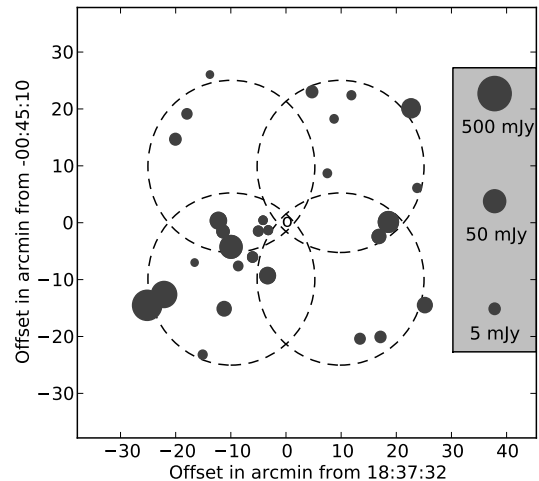}
         \caption{The pointing pattern used to search for in--beam calibrators.  The dashed lines show the 
         half--power point of the VLBA beam, an open circle represents the pulsar and the filled circles show 
         the background NVSS sources.  The circle diameter is proportional to the logarithm of the
         NVSS integrated flux density. 
         This example shows the field surrounding PSR J1837-0045.}
         \label{fig:inbeampointing}
\end{figure}

The first phase of the \psrpi\ project is a search for in--beam calibrators around potential
astrometric targets.  Figure~\ref{fig:inbeampointing} shows an example of the pointing layout used, with known 
background sources identified by circles.  In each case,
the observations were referenced to the nearest known suitable VLBA calibrator (defined here as R)
using the following scan sequence:

R-P1-P2-R-P3-P4-R-P1-P2-R-P3-P4-R

Two minute scans on each target pointing were used, for a total on--source time of four minutes per pointing.
Generally, five target pulsars with relatively small angular separations were grouped and observed sequentially in a single 
observation to minimise calibration overhead.
Correlation was performed using the DiFX software correlator \citep{deller11a} with one phase centre placed
on each known source.  Sources were extracted from the FIRST survey  \citep{becker95a} 
where available, and the NVSS survey \citep{condon98a} in areas not covered by FIRST.
Correlation was performed with a spectral resolution of 4 kHz to avoid bandwidth
smearing, but the output visibilities were averaged in time to 4 second resolution and in 
frequency to 1 MHz resolution.  

Currently, the in--beam search observations for \psrpi\ are nearly complete, with 77 hours of
observing time expended searching around 200 pulsars.  Over 7000 potential calibrators have been
imaged, with 530 detected in our VLBI maps at $>9\sigma$.  One or more satisfactory in--beam calibrators were found for 97\% of sources, with the remainder
largely being lost due to scattering in the Galactic plane.  
The detection rate was comparable for FIRST and NVSS sources; although the FIRST catalog is more compact on
average, it also contains objects fainter than our VLBI detection threshold, so this is not surprising.
These observations were optimized to locate sufficiently bright in--beam calibrators,
and so the raw results in terms of VLBI detection fractions should be interpreted with caution.

\section{Astrometric scheduling}   \label{sec:astrometry}
Of the sources with one or more sufficiently bright in--beam calibrator candidates, 110 were
selected for exploratory astrometric observations. The exploratory
phase is used to confirm the suitability of both the pulsar and the in--beam calibrators, since the 
pulsar flux density is often uncertain and the snapshot observations provide limited constraints on 
source morphology.   From these observations, 60 sources
will be selected for full astrometric observations encompassing 8 epochs over 1.5 years.  

\begin{figure*}[!ht]
         \centering
         \begin{tabular}{cc}
          \includegraphics[width=.44\textwidth]{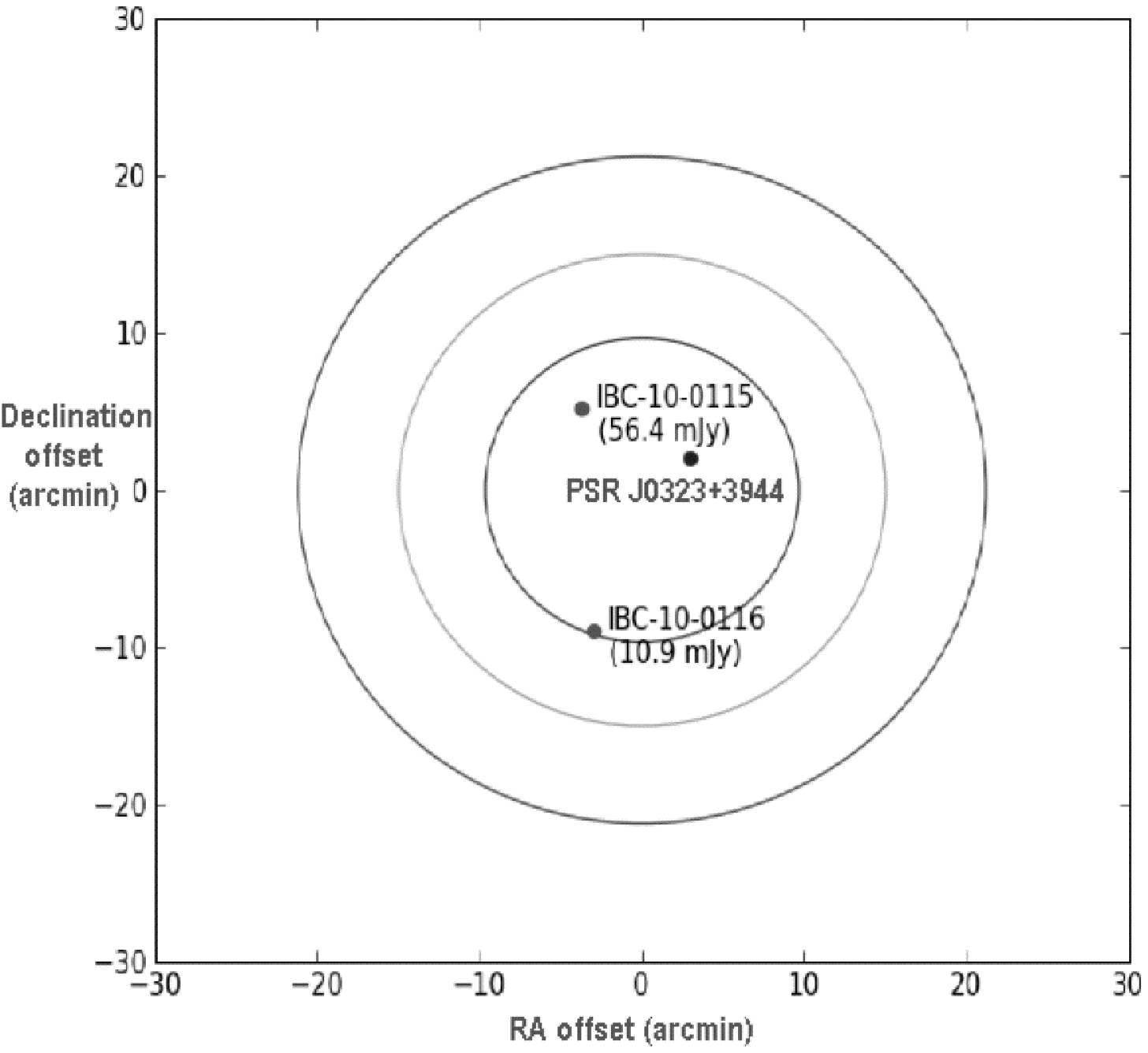} &
          \includegraphics[width=.4\textwidth]{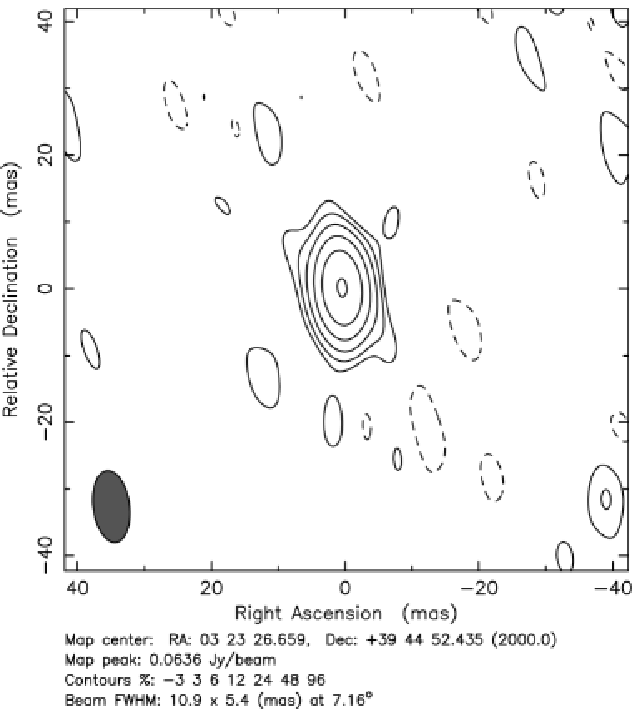} \\
          \includegraphics[width=.4\textwidth]{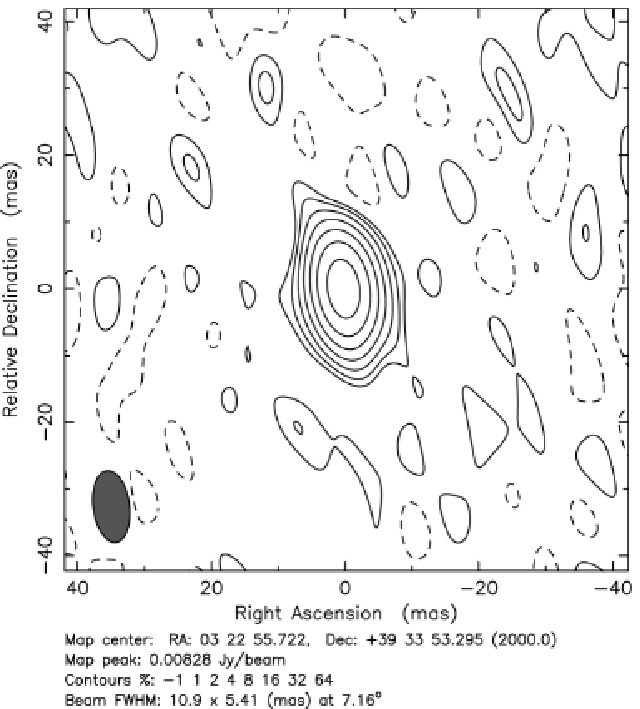} &
          \includegraphics[width=.4\textwidth]{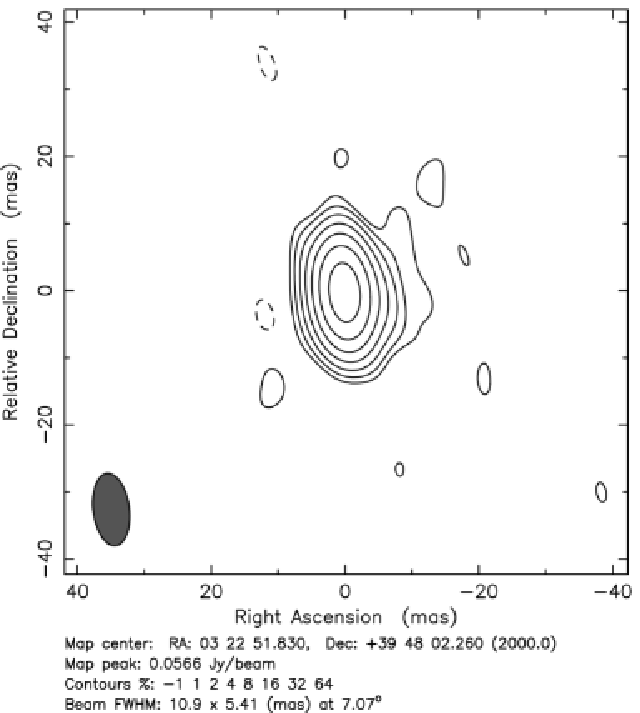} \\
          \end{tabular}
         \caption{Results from the first astrometric epoch on PSR J0323+3944.  Clockwise from top left: the pointing setup (concentric circles showing the 75\%, 50\% and 25\% response points of the beam), the pulsar detection from a gated correlation, the primary inbeam calibrator, and the secondary inbeam calibrator.}
         \label{fig:sampleexploratory}
\end{figure*}

For both exploratory and final astrometric observations, pulsars are grouped in pairs
to minimise calibration overhead and improve $uv$ coverage.  Each pulsar is observed
 in two groups of phase referenced scans, with 30 minutes
of on--source time per group.  The remainder of the time was spent slewing or on external calibrator sources. 
Matched filtering \citep{deller07a} is applied when correlating pulsar data, using pulsar ephemerides obtained from
pulsar timing.  Figure~\ref{fig:sampleexploratory} shows the pointing setup
and the VLBI detections of the source PSR J0323+3944 and its two in--beam calibrators.

Several possibilities exist for taking advantage of the fortunate case of multiple in--beam 
calibrator sources. The simplest is
to re--reduce the datasets using each calibrator independently \citep{chatterjee09a}.  This
provides robustness against calibrator variability, but cannot easily
take advantage of the partially independent results. A more sophisticated approach 
is to appropriately weight the calibrator visibilities and coherently sum them.  This averages over  
the partially independent atmospheric errors, and also reduces the random component of the solution error because of the improved
S/N.  Sources too faint to provide solutions in isolation can then be used in the sum, but each source must be
accurately modeled in order to be added coherently.  Furthermore, if any of the included sources
varies in position then the sum and the astrometric results may be adversely affected.

A third, and ideal, possibility is that three or
more calibrators would be used to solve for a calibration plane, rather than point, above each antenna
at regular intervals in the observation.  Whilst it is presently challenging to obtain sufficiently high signal to noise 
observations on three or more calibrators, it may be possible for a small number of \psrpi\ targets, and future
high--sensitivity, wide--field instruments such as the Square Kilometre Array (SKA) should use this approach  
routinely.  Further opportunities offered by multi--source astrometric datasets are discussed in Section~\ref{sec:multical}
below.

\section{\psrpi\ outcomes}   \label{sec:outcomes}
The completion of the first phase of \psrpi\ will bring about a six--fold increase in the number of pulsars 
with a parallax measured to an accuracy of 50\uas\ or better.  Combined with the small number of existing
accurate pulsar distance measures, this will give a moderately sized ($\sim$70 objects) pool of pulsars
with reliable distance measurements which can be used for unbiased investigations of the pulsar luminosity
and velocity functions. The significant increase in the 
number of accurate distance measures will also be used as part of the next major upgrade to the 
Galactic electron density distribution model, following NE2001 \citep{cordes02a}.  The addition of a large
number of new sightlines with accurate electron column density measurements will enable
a significant improvement to the model, both in terms of removing systematic biases in the large scale structure, 
and including many refinements in the form of small scale under-- and over--densities within several kpc of the 
solar system.

Finally, in a sample of 60 pulsars, it is expected that a significant number of unexpected and interesting 
results will be uncovered based on individual pulsars.  These could take the form of high velocity
pulsars \citep[e.g.,][]{chatterjee05a}, associations with supernova remnants \citep[e.g.,][]{thorsett02a},
revision to measurements of high--energy emission \citep[e.g.,][]{deller09b}, or breaking
degeneracies in pulsar timing models \citep[e.g.,][]{deller08b}.

\section{Novel applications of multi--source astrometric datasets}   \label{sec:multical}
Studying the structural evolution of calibrator sources at the \uas\ level
is challenging using absolute astrometry \citep{titov11a}, and is practically impossible for
faint sources.  However, source structure evolution is likely to become the limiting factor for 
extremely precise astrometry using future instruments such as the SKA, and so understanding its characteristics
is a topic of considerable importance. 
Relative astrometry offers the precision necessary to probe this regime,
but the use of a single calibrator--target pair leaves an ambiguity as to the source of
any variability.  Multiple--source datasets, however, offer an opportunity to break this degeneracy.

Overcoming the usual limitations of sensitivity and the ionosphere is difficult; discerning evolution
at the level of 10~$\mu$as or below would limit the angular separation to a maximum of 5' (which dictates the
use of sub--mJy targets) and require a S/N ratio of several hundred (meaning an image rms
of $\sim$10 $\mu$Jy must be obtained).  Although challenging, such observations are possible today using 
the European VLBI Network or the High Sensitivity Array.

Such an observing program could be used to monitor the source structure stability of a small
sample of ICRF sources, allowing better estimation of the level at which source structure evolution
contaminates the ICRF (by using the ICRF source as a calibrator and investigating the 
correlated motion of the other in--beam sources).  This could be undertaken commensally
with other astrometric projects using these calibrators.  However, it would be necessary to 
obtain a deep image of the field around the source with a lower resolution instrument  to first identify the
potential VLBI sources.

\section{Conclusions}   \label{sec:conclusions}
The \psrpi\ program, which will measure 60 pulsar parallaxes to an accuracy of better than 50 $\mu$as,
is underway with the VLBA.  A new capability of the DiFX software correlator has been used to greatly
speed up the identification of the in--beam calibrators necessary to reach this level of accuracy,
with over 7,000 potential targets imaged using less than 80 hours of observing time.  Once completed, the \psrpi\ program 
will provide the necessary information for a substantial improvement in the Galactic electron density distribution
model, and the improved distance information for the pulsar population (both directly for some pulsars and through
the model refinement for the remainder) will be used to update models of the pulsar velocity and luminosity
distributions.  Finally, the \psrpi\ dataset will provide opportunities to explore calibration techniques relevant
to future interferometers such as the SKA.

\bibliographystyle{apj}
\bibliography{deller_thesis}

\end{document}